\documentclass[aps,prb,twocolumn,showpacs,10pt,floatfix,longbibliography]{revtex4-2}
\usepackage{siunitx, tikz}
\usepackage{graphicx, subfigure,amsmath, amsthm, amssymb,  braket, mhchem}
\usepackage[T1]{fontenc}

\usepackage[bookmarks=true, colorlinks, citecolor=blue, linkcolor=blue, urlcolor=blue]{hyperref}
\usepackage{bm}
\usepackage{color}

\usepackage{ulem}

\begin{document}

\title{Evanescent-mode-assisted Klein tunneling in dual-gated bilayer graphene}


\author{Yupeng Huang$^{1}$}
\author{W.~Zeng$^2$}
\email{zeng@ujs.edu.cn}
\affiliation{$^{1}$School of Science, Jiangxi University of Science and Technology, Ganzhou 341000, China\\
$^{2}$Department of Physics, Jiangsu University, Zhenjiang 212013, China}


\date{\today}

\begin{abstract}
We theoretically investigate the electron tunneling in dual-gated bilayer graphene-based $n/p$ junctions. It is shown that a band gap is introduced by tuning the gate voltage, which modifies the pseudospin polarization and breaks anti-Klein tunneling at normal incidence. Specifically, when the pseudospin polarization vectors for the propagating and evanescent wave modes on the left and right regions of the junction are orthogonal, a revival of Klein tunneling is achieved. The Berry phase associated with Klein tunneling in dual-gated bilayer graphene is not limited to $\pi$ but varies with the junction parameters. Furthermore, the Klein tunneling is accompanied by a $\pi$ jump in the reflection phase around the normal incidence.
\end{abstract}


\maketitle

\section{Introduction}
Bilayer graphene (BLG), similar to its monolayer counterpart, has attracted great interest due to its unique physical properties. Bilayer graphene consists of two coupled monolayers, arranged such that one atom of the $B$ sublattice in the lower layer is directly coupled to an atom of the $A$ sublattice in the upper layer \cite{McCann_2013, MCCANN2007110, doi:10.1126/science.1130681}, where $A$ and $B$ denote the two inequivalent lattice sites in the monolayer honeycomb lattice. The other two sublattices, \textit{i.e.}, $A$ sublattice in the lower layer and the $B$ sublattice in the upper layer, do not have a counterpart, and sit exactly over the center of the hexagon on the other layer. Such a stacking arrangement is known as Bernal-stacked bilayer graphene \cite{McCann_2013}, where the charge carriers behave like massive Dirac fermions, providing platform for investigating a series of novel quantum properties, such as the quantum Hall effect \cite{novoselov2006unconventional, PhysRevB.78.033403, PhysRevB.87.121408, doi:10.1126/science.adj3742}, the spin-orbit coupling \cite{PhysRevB.85.115423, zhang2023enhanced, island2019spin} and the tunneling valley polarization \cite{PhysRevB.82.165409, PhysRevB.103.115436, PhysRevB.103.075414, PhysRevLett.130.266003, chen2020gate, Wu_2024}.

Klein tunneling is a remarkable transmission phenomenon in monolayer graphene (MLG), where a low-energy electron incident normally on a potential barrier can transmit through the barrier regardless of the barrier height, analogous to the relativistic quantum tunneling \cite{katsnelson2006chiral, ando1998berry, allain2011klein, Tudorovskiy_2012}. This effect stems from the pseudospin-momentum locking of electrons, where the reflection requires a reversal of the pseudospin, yet such pseudospin reversal is prohibited, thereby suppressing backscattering entirely. It is also attributed to the $\pi$-valued Berry phase associated with the pseudospin, as normal injection is accompanied by a $\pi$ jump in the reflection phase \cite{ando1998berry, PhysRevLett.101.156804, SHYTOV20091087, young2009quantum}. In BLG, the pseudospin rotates twice as fast as the momentum, resulting in a Berry phase of $2\pi$. Consequently, electrons incident normally on a potential barrier in BLG are reflected with unit probability, as the transmission demands pseudospin reversal. This phenomenon is termed as anti-Klein tunneling \cite{katsnelson2006chiral, PhysRevB.84.235432, PhysRevLett.132.146302, PhysRevB.109.115429}.

One notable feature of BLG is the ability to induce and modulate a band gap by tuning the interlayer potential difference using dual-split gates, where the pseudospin configuration is altered \cite{McCann_2013, PhysRevB.74.161403, oostinga2008gate, zhang2009direct}. Introducing a band gap enables the generation and control of valley polarization in electrons. Previous works predict that a steplike profile of the electrostatically defined interface separating regions of different pseudospin polarization would give rise to valley polarization \cite{PhysRevB.82.165409, PhysRevB.92.045417}. Recently, it has been demonstrated that the valley polarization may disappear by tuning potential step and band gap in a monopolar $n/n^\prime$ (or $p/p^\prime$) junction, indicating that a band gap is a necessary but insufficient condition for the valley polarization in dual-gated BLG \cite{Wu_2024}.

In addition, the band gap also significantly affects the anti-Klein tunneling. The perfect reflection at normal incidence can be reduced by inducing the band gap in BLG. A transition from the anti-Klein tunneling to the nearly perfect Klein tunneling is possible by modulating the Berry phase, which can be achieved via the band gap adjustment. It is assumed that the perfect Klein tunneling requires the Berry phase to be $\pi$ in both regions of a $n/p$ junction, and a maximum transmission probability of 0.87 at normal incidence has been observed in experiments \cite{https://doi.org/10.1002/pssr.201510180, PhysRevLett.113.116601, PhysRevLett.121.127706}. Over the past decade, both experimental and theoretical efforts have been devoted to this research. However, a clear theoretical description of this peculiar transition still remains to be established.

In this paper, we theoretically investigate the electron tunneling through $n/p$ junctions based on dual-gated BLG. By tuning the gate voltage, a band gap is introduced in the energy dispersion, which modulates the pseudospin of electrons in BLG. We demonstrate that Klein tunneling occurs in BLG when the pseudospin vector of the propagating (evanescent) wave in the $n-$doped region is orthogonal to that of the evanescent (propagating) mode in the $p-$doped region, and an exact analytical formula that describes the condition of Klein tunneling is presented. We further show that Klein tunneling is accompanied by a jump in the reflection phase by $\pi$ around the normal injection. In contrast, when the Klein tunneling is suppressed, the reflection phase exhibits a continuous variation near the normal incidence.

The remainder of the paper is organized as follows. The model Hamiltonian and the scattering approach are given in Sec.\ II. In Sec.\ III, the numerical results and discussions are presented. Finally, we briefly conclude in Sec.\ IV.

\section{Theoretical Model}
\begin{figure}[tbp]
	\centering
	\subfigure{
		\includegraphics[width=\linewidth]{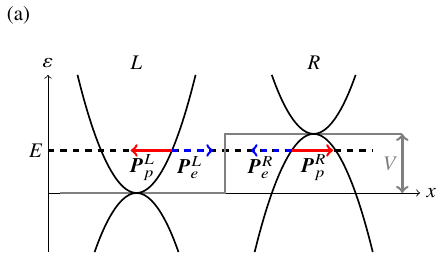}}
	\subfigure{
		\includegraphics[width=\linewidth]{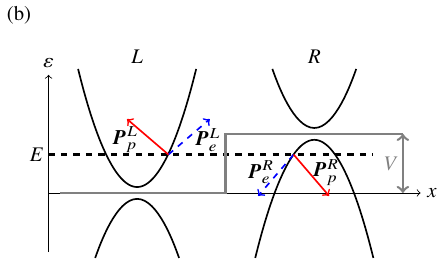}}
	\caption{\label{Fig:1}Schematic diagram depicting a sharp potential step at $x=0$ in the $n/p$ junction based on the dual-gated BLG. The dashed lines represent the Fermi level. The longitudinal direction of the junction is along the $x$ axis. The $n$-doped ($p$-doped) region is located at $x<0$ ($x>0$), which is labeled by $L$ ($R$). (a) The pseudospin polarization vectors of the normally incident electrons when the anti-Klein tunneling occurs. The red solid arrows represent the pseudospin polarization vectors of the propagating modes, and the blue dashed arrows represent the pseudospin polarization vectors of the evanescent modes. (b) The pseudospin polarization vectors of the normally incident electrons when the Klein tunneling occurs. The pseudospin polarization vectors for the propagating and evanescent states on the left and right sides of the $n/p$ junction are orthogonal.
	}
\end{figure}

We consider the two-dimensional $n/p$ junction based on BLG along the $x$ axis, where the $n$- and $p$-doped regions are located at $x<0$ and $x>0$ segments, respectively, as shown in Fig.\ \ref{Fig:1}. The low-energy Hamiltonian reads \cite{PhysRevLett.96.086805, PhysRevB.74.161403}
\begin{equation}
\begin{split}
\mathcal{H}_{\text{BLG}}&=\frac{v^2}{\gamma}\bm{\sigma}\cdot\left (-p_x^2+p_y^2,-2\xi p_xp_y,\frac{\gamma\Delta}{v^2}\right )+V(x),\label{Eq:1}
\end{split}
\end{equation}
which acts on the basis $(\psi_{B1}, \psi_{A2})^\text{T}$ with $A$ and $B$ denoting the two inequivalent sublattices of a monolayer graphene, and the indices 1 and 2 referring to the top and bottom layers of bilayer graphene, respectively. $p_{x(y)}=-i\hbar\partial_{x(y)}$ is the momentum operator in the $x(y)$ direction, $\bm\sigma=(\sigma_x,\sigma_y,\sigma_z)$ is the pseudospin Pauli matrix,
$\gamma\simeq\SI{390}{meV}$ is the interlayer nearest neighbor hopping energy \cite{PhysRevB.80.165406,Mucha-Kruczyński_2010}, $v\simeq 10^6\text{m/s}$, and $\xi=1(-1)$ corresponds to the $K$($K^\prime$) valley. For simplicity, we set $\hbar=v=\gamma=1$ throughout our calculation. $V(x)$ and $\Delta$ are the electrostatic potential and the pseudospin polarization term, respectively. Both terms are independently controlled by the top- and bottom-gate voltages \cite{PhysRevB.74.161403,PhysRevLett.99.216802}. We assume the electrostatic potential is smooth on the atomic scale but sharp on the length scale of the Fermi wavelength. In this case, the potential $V(x)$ can be expressed in a step model given as
\begin{equation}
	\begin{split}
		V(x)=\begin{cases}
			V_L, & x\leq0, \\
			V_R, & x>0.
		\end{cases}
	\end{split}\label{Eq:2}
\end{equation}

The scattering states $\psi(\bm r)$ can be obtained by solving the Schr$\ddot{\text{o}}$dinger equation
\begin{equation}
	\mathcal{H}_{\text{BLG}}\psi(\boldsymbol{r})=\varepsilon\psi(\boldsymbol{r}),\label{Eq:3}
\end{equation}
yielding a parabolic energy spectrum
\begin{equation}
	\varepsilon=\pm\sqrt{(k_x^2+k_y^2)^2+\Delta^2}+V(x),\label{Eq:4}
\end{equation}
where $\pm$ corresponds to the conduction and valence bands, respectively. The pseudospin polarization term induces a band gap valued $2\Delta$ in the dispersion. For a given energy $E$ and conserved transverse wave vector $k_y$, $k_x$ has two solutions corresponding to propagating modes ($\psi_{j,p}$) and evanescent modes ($\psi_{j,e}$), which are given by
\begin{equation}
	\begin{split}
		\psi_{j,p}^\pm=&\frac{1}{N_j}\begin{pmatrix}
			E-V_j+\Delta \\
			-(\pm s_jk_j+i\xi k_y)^2
		\end{pmatrix}e^{\pm is_jk_jx}, \\
		\psi_{j,e}^\pm=&\begin{pmatrix}
			E-V_j+\Delta \\
			(\pm\kappa_j+\xi k_y)^2
		\end{pmatrix}e^{\mp\kappa_jx},
	\end{split}\label{Eq:5}
\end{equation}
respectively, with
\begin{equation}
	\begin{split}
		s_j&=\mathrm{sgn}(E-V_j), \\
        N_j&=\sqrt{2(E-V_j)(E-V_j+\Delta)}, \\
		k_j&=\{\sqrt{(E-V_j)^2-\Delta^2}-k_y^2\}^{1/2}, \\
		\kappa_j&=\{\sqrt{(E-V_j)^2-\Delta^2}+k_y^2\}^{1/2}. \\
	\end{split}\label{Eq:6}
\end{equation}
Here the superscript $+(-)$ in Eq.\ (\ref{Eq:5}) represents the right (left) propagating direction. The subscripts $j=L,R$ in Eqs.\ (\ref{Eq:5}-\ref{Eq:6}) refer to the left ($x\leq0$) and the right ($x>0$) regions of the junction, respectively. The transverse wave vector $k_y$ is conserved because of the translational invariance along the $y$ direction, and we omit $e^{ik_yy}$ term in Eq.\ (\ref{Eq:5}). 

We choose the parameter set $V_L=0$ and $V_R=V>E$ in our model, \textit{i.e.} $s_L=-s_R=1$, resulting in an $n/p$ junction as shown in Fig.\ \ref{Fig:1}. The wave functions in the junction is a superposition of propagating and evanescent waves
\begin{equation}
	\Psi(x)=\begin{cases}
		\psi_{L,p}^++r\psi_{L,p}^-+r_e\psi_{L,e}^-, & x\leq0, \\
		t\psi_{R,p}^++t_e\psi_{R,e}^+, & x>0,
	\end{cases}\label{Eq:7}
\end{equation}
where $r$ ($t$) is the reflection (transmission) amplitude for the propagating waves, and $r_e$ ($t_e$) is the reflection (transmission) amplitude for the evanescent waves. The scattering coefficients are solved by matching the wave functions at the boundary, which involves the continuity of $\Psi(x)$ and $d\Psi(x)/dx$ at $x=0$, corresponding to the flux conservation. As a result, the reflection and transmission probability are given by $R=|r|^2$ and $T=|t|^2$, respectively.

\section{Results and Discussion}

\begin{figure}[tbp]
	\centering
	\includegraphics[width=\linewidth]{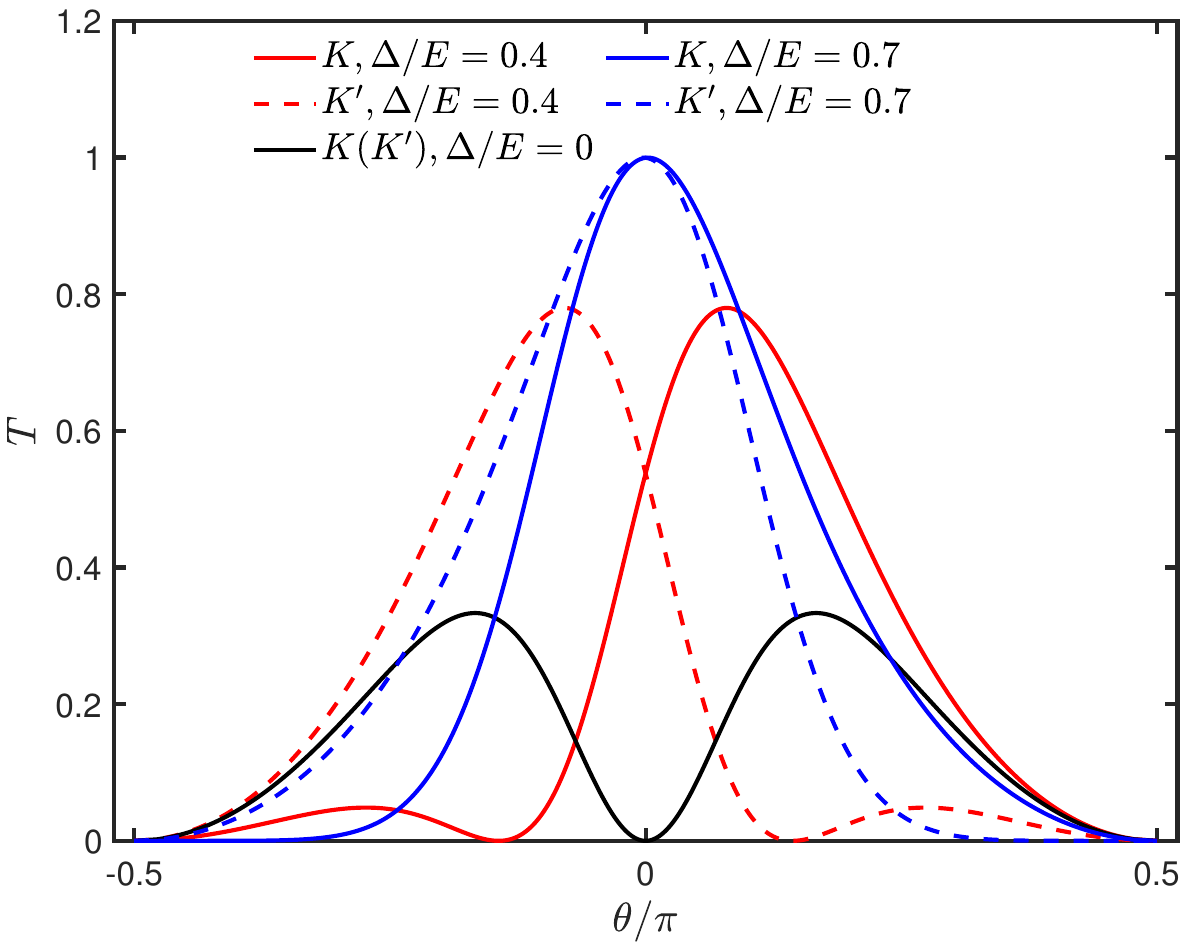}
	\caption{\label{Fig:2}The transmission probability $T$ versus the incident angle $\theta$ at $V/E=2$ for different $\Delta$. The solid and dashed lines represent the $K$ and $K'$ valleys, respectively.}
\end{figure}

We first consider the transmission probability $T$ as a function of incidence angle $\theta=\arctan(k_y/k_L)$ for a given incident energy $E$ through a sharp potential of height $V$. The numerical results at $V/E=2$ are shown in Fig.\ \ref{Fig:2}. For $\Delta=0$, a perfect reflection at normal incidence is observed; see Fig.\ \ref{Fig:2} (black line). This phenomenon is known as anti-Klein tunneling \cite{katsnelson2006chiral}, which can be understood in terms of pseudospin conservation, as shown in Fig.\ \ref{Fig:1}(a), where the pseudospin polarization vector of the transmission electron is anti-parallel to that of the incident one, and thus the normally transmitted electron is completely forbidden.

By applying a different potential to the two layers of the BLG, the inversion symmetry is broken due to the presence of the finite bandgap $\Delta$. The transmission probabilities for $K$ and $K^\prime$ valleys differ significantly at oblique incidence ($\theta\neq0$), but exhibit a valley-exchange symmetry, \textit{i.e.}, $T_K(\theta)=T_{K^\prime}(-\theta)$, as shown in Fig.\ \ref{Fig:2}. Due to the time-reversal symmetry, the total transmission probability [$T_K(\theta)+T_{K'}(\theta)$] is symmetric with respect to the incidence angle. We note that this valley-dependent skew tunneling may result in the tunneling valley Hall effect in our $n/p$ junction model \cite{PhysRevB.110.205406, PhysRevB.111.075418, PhysRevLett.131.246301}. For normally incident electrons ($\theta=0$), the transmission probabilities for different valleys are the same, which is attributed to the fact that the Hamiltonian becomes valley-independent at $k_y=0$. The presence of the bandgap $\Delta$ has a significant impact on the normally tunneling properties. Remarkably, the anti-Klein tunneling is broken with a finite transmission probability. At $V/E=2$, the normally incident transmission probability [$T(\theta=0)$] increases from $0$ to $1$ as $\Delta$ rises from $0$ to $0.7E$, indicating the transition from anti-Klein tunneling to Klein tunneling.

\begin{figure}[tbp]
	\centering
	\subfigure{
		\includegraphics[width=\linewidth]{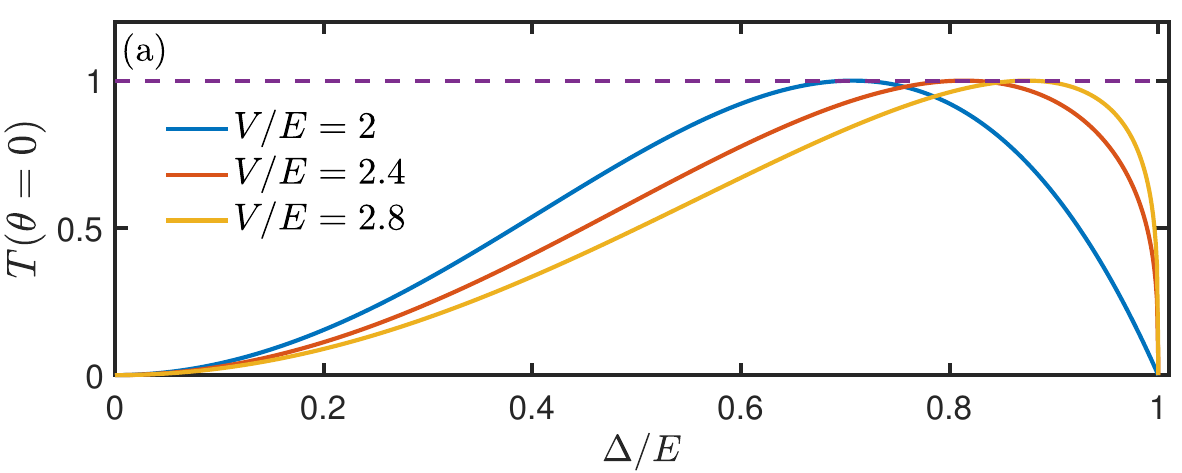}}
	\subfigure{
		\includegraphics[width=\linewidth]{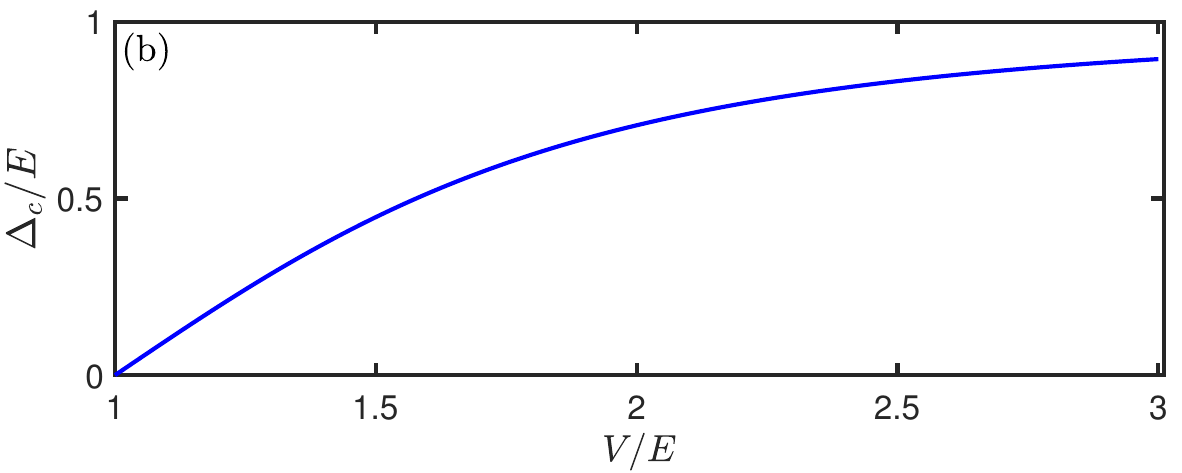}}
	\caption{\label{Fig:3}(a) The normally incident transmission probability $T(\theta=0)$ versus the band gap $\Delta$ for different values of potential height $V$. (b) The critical value $\Delta_c$ versus the potential height $V/E$.}
\end{figure}

This transition predicted in our model also persists as the potential height $V$ varies. The transmission probability for the normally incident electrons $T(\theta=0)$ as a function of the bandgap $\Delta$ at different $V$ is shown in Fig.\ \ref{Fig:3}(a). It is shown that the transition point $\Delta_c$, satisfying 
\begin{align}
	T(\theta=0)|_{\Delta=\Delta_c}=1, \label{eq:tt}
\end{align}
exhibits a systematic shift with the potential energy, indicating that the tunneling characteristics are strongly influenced by the electrostatic environment. The current conservation requires $T+R=1$, which indicates that the equivalent form of Eq.\ (\ref{eq:tt}) can be expressed as $R=|r|^2=0$. Consequently, with the help of
\begin{equation}
	\begin{split}
		&r(\theta=0)\\
		=&\frac{\sqrt{a_La_R}(e^{2\vartheta_L}+e^{2\vartheta_R})-2e^{\vartheta_L+\vartheta_R}(a_L+a_R+\sqrt{a_La_R})}{i\sqrt{a_La_R}(e^{2\vartheta_L}+e^{2\vartheta_R})+2e^{\vartheta_L+\vartheta_R}(a_L-a_R+i\sqrt{a_La_R})},\label{Eq:8}
	\end{split}
\end{equation}
where $a_j=\sqrt{(E-V_j)^2-\Delta^2}$ and $\vartheta_j=\mathrm{arcsinh}(s_j\Delta/a_j)$ with $j=L,R$, the transition point for the Klein tunneling is given by
\begin{equation} 
	\Delta_c=E\cos\left [\arctan\left (\dfrac{E}{V-E}\right  )\right ]. \label{Eq:9}
\end{equation}
The dependence of the transition point $\Delta_c$ on potential height $V/E$ is shown in Fig.\ \ref{Fig:3}(b). As can be seen, Eq.\ (\ref{Eq:9}) reveals that the transition point rises monotonically as $V$ increases, which is in agreement with the numerical results presented in Fig.\ \ref{Fig:3}(a). Moreover, according to Eq.\ (\ref{Eq:9}), the Klein tunneling occurs at $\Delta_c=E/\sqrt{2}\simeq0.7E$ for $V/E=2$, consistent with the numerical results shown in Fig.\ \ref{Fig:2}. 

Next, we further examine the relation obeyed by the pseudospin structure at the transition point. The pseudospin polarization vectors for the propagating and evanescent electron states at normal incidence are given by
\begin{equation} 
	\begin{split}
	\bm P_{p}^j&=\frac{\bra{\psi_{j,p}}\bm\sigma\ket{\psi_{j,p}}}{\braket{\psi_{j,p}|\psi_{j,p}}}=\left (\frac{-a_j}{E-V_j}, 0, \frac{\Delta}{E-V_j}\right ), \\
		\bm P_{e}^j&=\frac{\bra{\psi_{j,e}}\bm\sigma\ket{\psi_{j,e}}}{\braket{\psi_{j,e}|\psi_{j,e}}}=\left (\frac{a_j}{E-V_j}, 0, \frac{\Delta}{E-V_j}\right ),\label{Eq:10}
	\end{split}
\end{equation}
respectively. Here, the pseudospins of forward and backward states are identical, thus we omit the superscript $\pm$ in Eq.\ (\ref{Eq:10}). At the transition point $\Delta=\Delta_c$, one finds that the pseudospin polarization vectors satisfy 
\begin{align}
	\bm P_{p/e}^L\cdot \bm P_{e/p}^R=0,\label{eq:oo}
\end{align}
indicating that the Klein tunneling occurs precisely when the pseudospin vectors for the propagating and evanescent states on the left and right sides of the $n/p$ junction are orthogonal. Fig.\ \ref{Fig:1}(b) depicts the pseudospin structure when the Klein tunneling occurs, which clearly illustrates the orthogonality relation specified by Eq.\ (\ref{eq:oo}). We emphasize that although Eq.\ (\ref{eq:oo}) is equivalent to Eq.\ (\ref{Eq:9}), it is more fundamental, as it reveals the role of evanescent modes in the Klein process of BLG. In contrast, such Klein tunneling, mediated by evanescent modes, is not present in MLG.

In MLG, the Klein tunneling is typically attributed to the pseudospin conservation. The backward and forward propagating electrons carry opposite pseudospins as dictated by the $\bm k\cdot\bm\sigma$ Dirac-type Hamiltonian, leading to the absence of the backreflection. This unique pseudospin-momentum locking nature in MLG requires that the polarization vectors of the scattering states mapped onto the Bloch sphere must lie on the equator, corresponding to a $\pi$ pseudospin winding number, \textit{i.e.}, a $\pi$ Berry phase.

However, this argument does not hold in gapped BLG because the evanescent modes participate in the Klein tunneling process. As we have already illustrated, for $V/E=2$ and $\Delta/E=1/\sqrt{2}$, the Klein tunneling occurs even when the incident and transmitted states have opposite pseudospin polarizations [denoted by $\bm{P}^L_p$ and $\bm{P}^R_p$, respectively, satisfying $\bm{P}^L_p=-\bm{P}^R_p=(-1,0,1)/\sqrt{2}$)]. 

\begin{figure}[tbp]
	\centering
	\subfigure{
		\includegraphics[width=\linewidth]{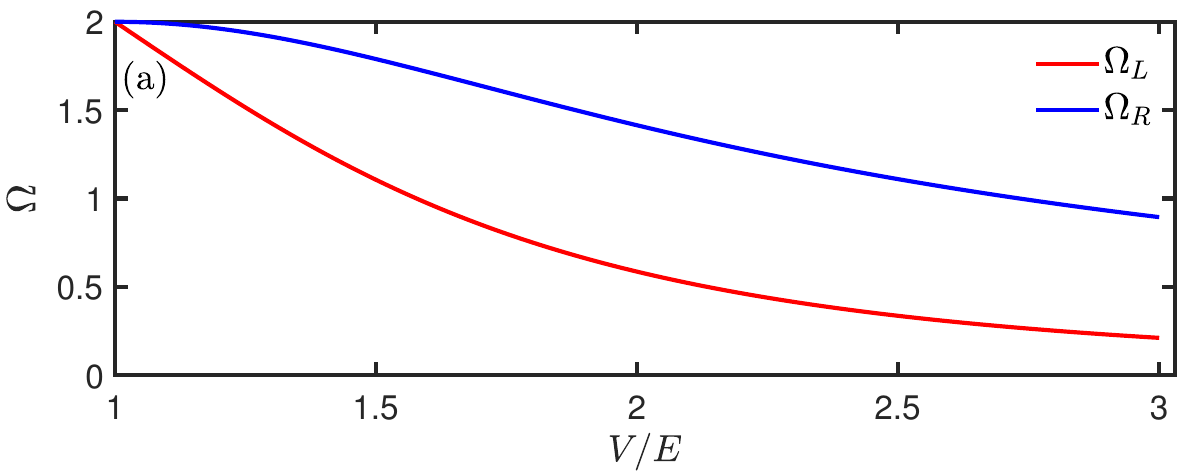}}
    \subfigure{
		\includegraphics[width=\linewidth]{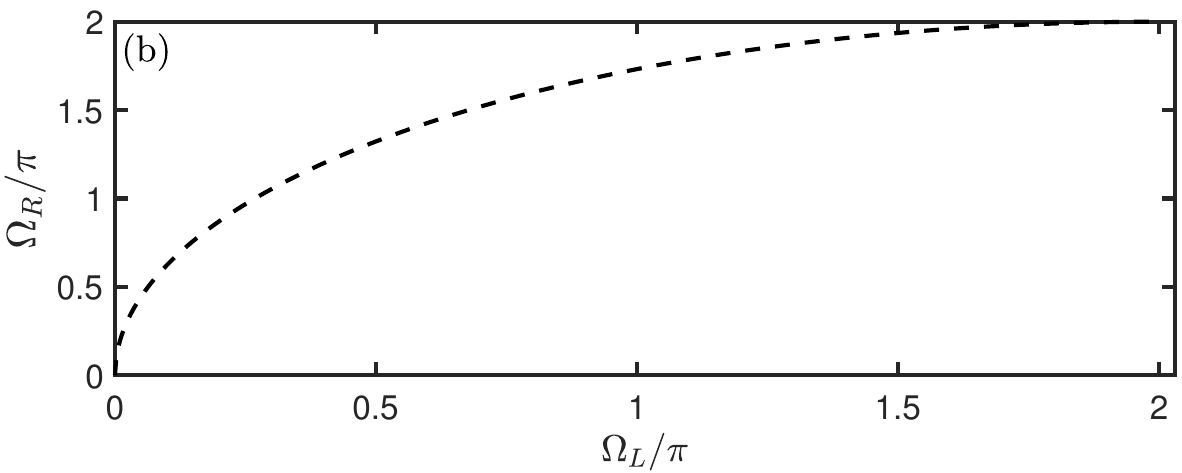}}
	\caption{\label{Fig:4}(a) The Berry phases in the two regions of a $n/p$ junction as functions of $V/E$ when Klein tunneling occurs. (b) Contour plot of the relation between $\Omega_L$ and $\Omega_R$ associated with Klein tunneling.}
\end{figure}

Moreover, in contrast to MLG, the Berry phase associated with Klein tunneling in BLG is not restricted to $\pi$, but instead varies with the junction parameters. The Berry phase can be obtained by $\Omega=\oint_\mathcal{C}d\bm{k}\bm\cdot\bm{A}$, where $\mathcal{C}$ is the closed path encircling the degeneracy point in the momentum space and $\bm{A}=i\langle\psi_{\bm{k}}|\nabla_{\bm{k}}|\psi_{\bm{k}}\rangle=\frac{2\nabla_{\bm{k}}\theta}{1+e^{2\phi_j}}$ with $j=L,R$ is the Berry connection. When the Klein tunneling occurs, one finds that the Berry phase in the left $n$-doped region is given by $\Omega_L=4\pi\sin^2[\frac{1}{2}\arctan(\frac{E}{V-E})]$, whereas in the right $p$-doped region it is given by $\Omega_R=\mod{(4\pi\cos^2[\frac{1}{2}\arctan(\frac{V-E}{E})],2\pi)}$, resulting in the following relation between the Berry phases across the $n/p$ junction:
\begin{equation}
	\arcsin\sqrt{\frac{\Omega_L}{4\pi}}+\arccos\sqrt{\frac{\Omega_R}{4\pi}}=\frac{\pi}{4}.\label{Eq:13}
\end{equation}
When Klein tunneling occurs, the Berry phases in the two regions of the junction decrease monotonically as $V/E$ increases, as illustrated in Fig.\ \ref{Fig:4}(a). Both $\Omega_L$ and $\Omega_R$ vary between $2\pi$ and $0$, yet differ from each other, satisfying a relation depicted by Fig.\ \ref{Fig:4}(b). Specifically, when $\Omega_L=\pi$, $\Omega_R=\sqrt{3}\pi$ is required to fulfill the condition for Klein tunneling.

\begin{figure}[tbp]
	\centering
	\includegraphics[width=\linewidth]{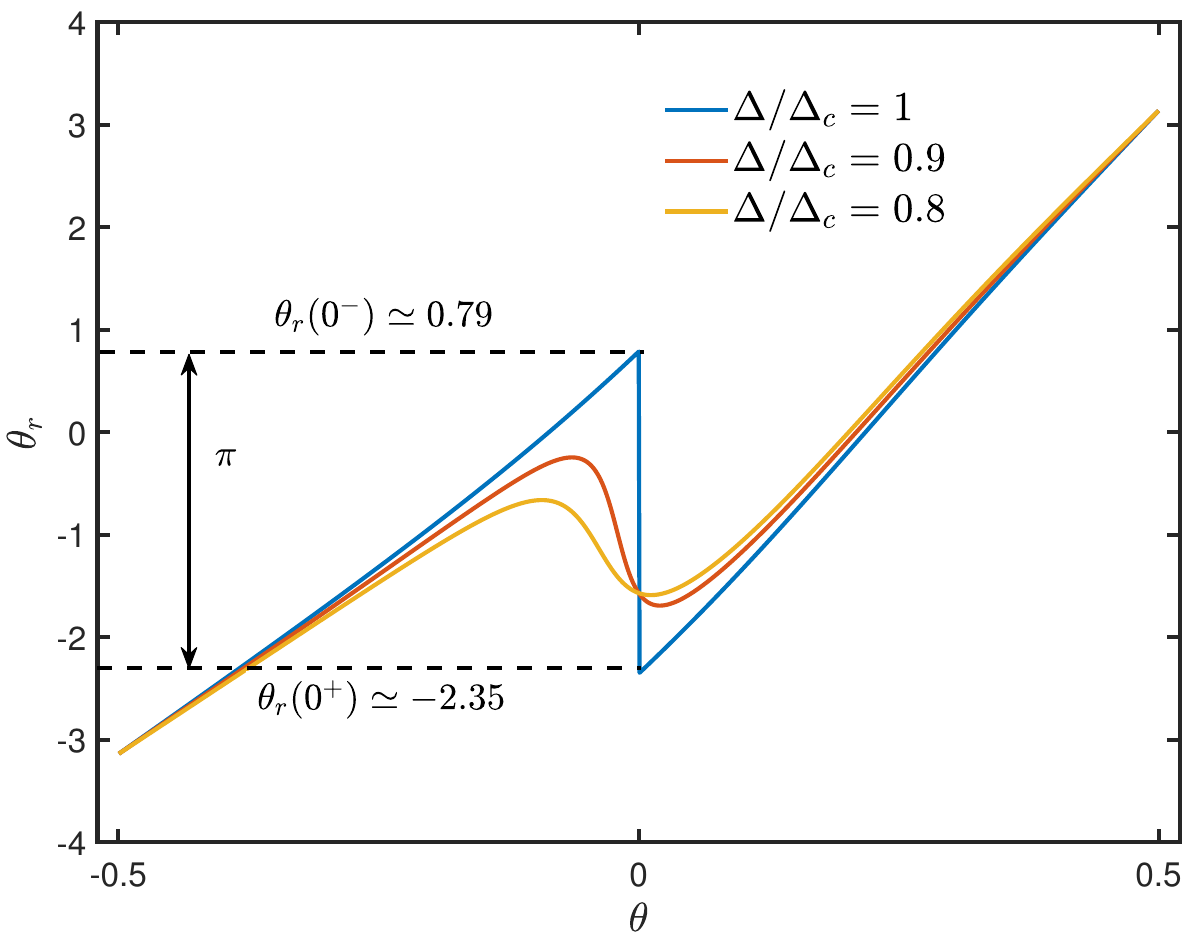}
	\caption{\label{Fig:5}Dependence of the backreflection phase $\theta_r$ on the incident angle $\theta$ for the $K^\prime$ valley at $V/E=2$. A $\pi$ phase jump around $\theta=0$ occurs when the Klein tunneling is present (blue line), but disappears when the Klein tunneling is suppressed (red and orange lines).
    }
\end{figure}

Similar to Klein tunneling in MLG \cite{PhysRevLett.101.156804}, Klein tunneling in BLG is also accompanied by a $\pi$ jump in the reflection phase at normal incidence \cite{PhysRevB.84.235432}. When the Klein tunneling occurs, one finds that the reflection amplitude can be expressed as 
\begin{align}
 	r\simeq r_0\cdot\theta+\mathcal{O}(\theta^2), \label{Eq:14}
 \end{align}
 where 
 \begin{align}
 	r_0=\frac{2\xi(1+i)\sinh(\vartheta_1-\vartheta_2)\sqrt{a_1a_2}}{(\sqrt{a_1}+\sqrt{a_2})^2-i(a_1-a_2)}
 \end{align}
 is a $\theta$-independent parameter. Consequently, the backreflection phase exhibits a discontinuity near $\theta=0$, with the magnitude of the jump equal to
 \begin{align}
   	\Delta\theta_r&=\arg[r(0^+)]-\arg[r(0^-)]\nonumber\\
    &=\arg[r_0\cdot|\eta|]-\arg[-r_0\cdot|\eta|],\quad \eta\rightarrow0^+\nonumber\\
    &=-\pi. \label{eq:pj}
 \end{align}  
 The backreflection phase $\theta_r=\arg(r)$ as a function of the incident angle $\theta$ is numerically calculated and is presented in Fig.\ \ref{Fig:5}. The $\pi$ phase jump predicted in Eq.\ (\ref{eq:pj}) is observed when the Klein tunneling occurs ($\Delta=\Delta_c$), as shown by the blue line in Fig.\ \ref{Fig:5}. However, this phase jump is absent when the Klein tunneling is suppressed. $\theta_r$ exhibits a continuous variation in the vicinity of $\theta=0$, as shown by the red and orange lines in Fig.\ \ref{Fig:5}, where the band gaps are chosen as $\Delta=0.9\Delta_c$ and $\Delta=0.8\Delta_c$, respectively.

\section{Conclusion}
To conclude, we theoretically investigate the transition from anti-Klein to Klein tunneling in dual-gated BLG $n/p$ junctions. We find that Klein tunneling occurs when the pseudospin vectors for the propagating and evanescent states on the left and right sides of the junction are orthogonal, this condition can be precisely described by Eq.\ \ref{eq:oo}. Furthermore, the occurrence of Klein tunneling even when the pseudospin vectors of propagating waves in the two regions are anti-parallel can be explained by the participation of evanescent waves in the tunneling process, which indicates that Klein tunneling in BLG arises from the combined effect of pseudospin configuration and evanescent waves. Moreover, the Berry phase associated with Klein tunneling varies between $2\pi$ and $0$ as the junction parameter $V/E$ increases. Additionally, a $\pi$ jump in reflection phase is accompanied around the normal injection, an effect that is absent when the Klein tunneling is suppressed.

\begin{acknowledgments}
	
	This work is supported by the Jiangxi provincial Department of Science and Technology under Grant No. 20252BEJ730189, and Jiangxi University of Science and Technology under Grant No. jxust-73.
\end{acknowledgments}

\bibliography{ref}

\end{document}